\documentclass[a4paper,11pt]{article}
\usepackage{pos}
\usepackage{caption}
\newcommand\apjs{{ApJS}}%         % Astrophysical Journal, Supplement  
\title{Search for multi-messenger events during LIGO/Virgo era}
\author*[a]{Tosta e Melo, I.}
\author[b]{LVK collaboration}

\affiliation[a]{INFN-LNS,\\
  Via S. Sofia 62, Catania, Italy}

\affiliation[b]{The LIGO Scientific Collaboration, the Virgo Collaboration, the KAGRA Collaboration\\
}

\emailAdd{tostaemelo@lns.infn.it}
\abstract{Multi-messenger astronomy is a vast and expanding field as electromagnetic (EM) observations are no longer the only way of exploring the Universe. 
Due to the new messengers, astrophysical events with both gravitational waves (GWs) and EM emission are no longer a dream of the astronomical community. 
A breakthrough for GW multi-messenger astronomy came when the LIGO-Virgo network detected a GW signal of two low-mass compact objects consistent with a binary neutron star (BNS, GW170817), an event that generated a short gamma-ray burst (GRB), and a kilonova. While GW170817 represents the testimony to BNS mergers being the progenitor of at least some GRBs, a wide range of highly energetic astrophysical phenomena is expected to be accompanied by the emission of GWs and photons. Here we present an unmodelled method to search for GWs having gamma and radio counterparts, using the LIGO/Virgo data and observations of partner telescopes. We also discuss the most recent results of the unmodelled coherent searches targeting astrophysical events during the first part of the LIGO-Virgo third observing run (O3a): 105 GRBs detected by the Fermi and Swift satellites. 
}

\FullConference{%
  EPS-HEP Conference 2021\\
  26-30 July, 2021\\
  Online conference
}

\begin{document}
\maketitle

\section{Introduction}

Nowadays, there is a tremendous development in multi-messenger astronomy as the EM observations are no longer the only way of exploring the Universe, especially after GW170817 \cite{abbott2017a,abbott2017b}.
This field had its dawn when new astronomical messengers of non-electromagnetic origin were observed: neutrinos from the Sun \cite{hirata} and from core-collapse supernova SN 1987A \cite{woosley}, high-energy neutrinos of extraterrestrial origin \cite{icecube}, and most recently the detection of GWs in 2015 (GW150914; \cite{abbott2016a}). 
A wide range of highly energetic transient (or burst) astrophysical phenomena is expected to be accompanied by the emission of GWs and photons, lasting from milliseconds to several seconds within the LVK (LIGO, Virgo and KAGRA) instruments frequency band \cite{ligocollab}. 

GRBs are among the plausible sources for the emission of both gamma-rays and GWs.
They are very strong pulses of gamma-rays from 10 ms to 1000 s and are considered the most energetic astrophysical events observed in the EM band.  
Historically they are divided into two classes based on their duration:
“short” and “long” \cite{zhang}. Long GRBs have a duration greater than 2 s, a softer spectrum, and their origin is related to the core collapse of rapidly rotating massive stars \cite{mosta}. 
Short GRBs have a duration of less than 2 s and a harder spectrum. Their progenitors are coalescing compact objects such as BNS or neutron star (NS)–black hole (BH) binary systems \cite{berger,nakar,abbott2017a,abbott2017b}. 
%After the detection of GW170817 and its $\gamma$ counterpart, GRB 170817A \cite{abbott2017a,abbott2017b} such a hypothesis was confirmed that compact binary mergers can produce short GRBs.
%In addition, compact binaries merger allow us to measure the masses and spins of the binary components \cite{hannam} and to constrain their relative merger rates as well as to place constraints on the beaming angle and the NS equation of state \cite{pannarale}.

Other possible interesting sources of GWs and EM counterparts are fast radio bursts (FRBs).
FRBs are bright, highly dispersed radio millisecond duration transients discovered over a decade ago \cite{lorimer}.
The sources of FRBs are still unknown and two separate populations are known to exist: repeaters and non-repeaters \cite{petroff}.
GW models for non-repeating FRBs include all types of compact binary coalescence (CBC) and unknown burst sources detectable by LVK instruments, although not all FRBs are likely to result in detection at current detectable radio frequencies.
%The easiest observable transient of GW signature is CBC, specifically NS-BH and BNs pairs, which may create at least some fraction of FRBs \cite{keane}.
%This may be due to the pulsarlike radio emission, either from the reactivation of the dormant pulsar emission in one of the NSs before the merger or by a hypermassive NS, which may occur before the collapse to a BH \cite{pshirkov}.
%Transient GW emission related to FRBs can occur when temporary deformation of a rapidly rotating NS creates a quadrupolar moment result of asteroseismics phenomena \cite{franco}.
%Also, NSs with high magnetization, known as magnetars, may occasionally and unpredictably give short flares which have been observed alongside quasi-periodic oscillations \cite{israel}.

%A targeted search for GWs in sky and time coincidence with a chosen trigger enhances our potential of achieving such joint detections. 
In this paper, we present the results for the targeted GW follow-up of GRBs reported by the Fermi \cite{meegan} and Swift \cite{barthelmy} satellites during the first part of the third observing run of Advanced LIGO \cite{advligo} and Advanced Virgo \cite{advvirgo} (O3a).
We processed 105 GRBs with a search for generic GW transients (X-Pipeline; \cite{sutton}). %Finally, present our concluding remarks.

\section{GRB sample}

Our GRB sample contains events by the Gamma-ray Coordinates Network (GCN) system\footnote{GCN Circulars Archive: http://gcn.gsfc.nasa.gov/gcn3archive.html.}, the Swift/BAT catalog\footnote{Swift/BAT Gamma-Ray Burst Catalog: http://swift.gsfc.nasa.gov/results/batgrbcat/}, the online Swift GRBs Archive\footnote{Swift GRB Archive: http://swift.gsfc.nasa.gov/archive/grb table/} and the Fermi GBM catalog\footnote{FERMIGBRST - Fermi GBM Burst Catalog: https://heasarc.gsfc.nasa.gov/W3Browse/fermi/fermigbrst.html.}. The GCN notices provide a set of 141 GRBs during the O3a data-taking period (2019 April 1 – 2019 October 1).

As discussed earlier, GRBs have two main classes based on their T$_{90}$ duration, the time interval over which 90$\%$ of the total background-subtracted photon counts are observed. 
In our analysis, GRBs are labelled as short when T$_{90}$ + |$\delta$T$_{90}$| is < 2 s. When T$_{90}$ - |$\delta$T$_{90}$| is > 4 s the GRBs are labeled as long, and the rest are labelled as ambiguous. Following this classification, we ended up with 20 short GRBs, 108 long GRBs, and 13 ambiguous GRBs. 

The unmodelled search for generic transients is applied to GRBs of all classifications. A modelled search was also conducted for the same sample of GRBs during O3a as well as in previous searches \cite{abbott2017c,abbott2019a,abbott2021}. 
GW searches for GRBs generally adopt a modelled search for the short duration bursts based on the assumption that coalescing compact binaries are the main progenitors for this class, but are unlikely for the long duration GRBs, that have been associated with supernovae. The unmodelled search is applied to both classes of GRB and requires a minimum of coincident data from at least two GW detectors around the time of a GRB trigger to assess the significance of a GW candidate with sub-percent level accuracy.
After applying this requirement to all GRBs, we ended up with 105 GRBs out of 141 to be analyzed during O3a. 
%From all 141 Fermi and Swift GRBs in our sample the vast majority do not have redshift measurements.

\section{Search Methodology}

We perform an analysis technique to extract the GW signal from the interferometers data that uses a coherent data stream and the information related to the time and sky position of an external trigger. 
It has been demonstrated \cite{was} that a triggered search can be up to 20$\%$ more sensitive than an all-sky search on the same data, depending on the accuracy of the sky position measurement, better for the Swift GRBs and less accurate for the Fermi-GBM ones. 
In order to perform such a search, we used a high-performing algorithm called X-Pipeline \cite{sutton}.

\subsection{Unmodelled search for generic transients}

X-pipeline is designed to search for GW bursts associated with an external astrophysical trigger. 
It looks for excess power that is coherent across the network of GW detectors and consistent with the sky localization and time window for each GRB. 
The sky localization is given by Fermi and Swift catalogs. The time window is defined by the GRB trigger time t$_0$ and the range of possible time delays between the GRB emission and the associated GW emission. 
We use an on-source search time window that begins 600 s before the GRB trigger time and ends 60 s after it or at T$_{90}$ if is greater than 60 s.
%as the on-source window to search for a GW signal. 
We also restrict our search to the most sensitive band of the GW detectors, i.e., 20–500 Hz, since GW energy necessary to produce detectable signals for GRBs are expected at low rather than high frequencies \cite{abbott2019a}.

X-pipeline coherently combines data from all detectors and produces time-frequency maps of the GW data stream. Clusters with an excess of energy in these maps are marked as events and then ranked according to a detection statistic based on their energy. 
These events pass through coherent consistency tests and the surviving event with the largest ranking statistic is marked as the best candidate for a GW detection, referred to as the “loudest event".
The significance of the loudest event is evaluated by comparing the signal-to-noise ratio (SNR) of the trigger within the on-source window to the SNR of the loudest triggers in the off-source trials.
%which require at least $\sim1.5$ hours of coincident data from at least two detectors around the GRB time. %This window is small enough to select data where the detectors should be in a similar state of operation as during the on-source interval, and large enough so that through artificial time-shifting.

We quantify the sensitivity of the generic transient search by injecting simulated signals into off-source data and recovering them. 
The simulated waveforms are chosen to cover different expected models such as: (i) BNS and NS-BH systems in form of inspiral signals, (ii) stellar collapses represented by circular sine-Gaussians (CSG) with central frequencies of 70, 100, 150, and 300 Hz, and (iii) accretion disk instability models represented by ADI waveforms as described in \cite{abbott2017c}.

\section{Results}

A search for GWs associated with GRBs detected by the Fermi and Swift satellites during the first part of the third observing run of Advanced LIGO and Advanced Virgo (2019 April 1 – 2019 October 1) was performed using X-pipeline on a total of 105 GRBs. 

We found no GW signal associated with any GRB in our sample. 
The evidence for the GRB and the GW coincidence is provided by estimating the probability of such signals to be detected in spatial and temporal agreement. 
The GW candidate event is defined as the loudest trigger found in the on-source window which passes all quality and coherent consistency tests.
The p-value associated with a GW candidate is the probability of the loudest event to be due to background fluctuations only. 
To select significant events that deserve a more detailed follow-up, we set a priori a threshold of p-value at 1$\%$ taking into account the trial factor of analyzing multiple GRBs. 
The cumulative distributions of p-values reported by the unmodelled search is given in Figure 1 (from \cite{abbott2021}), showing that all triggers fall in the expected background distribution under the no-signal hypothesis represented by the dashed line. 
Its 2$\sigma$ limits are indicated by the two dotted lines. 
The most significant event found in our analysis had p-value $5.5 \times 10^{-3}$ for the GRB 190804058. 
Such result is consistent with the estimated GW–GRB joint detection rate with Fermi-GBM of 0.07–1.80 per year reported in \cite{abbott2019b}. 

We placed limits on GW emission as we have found no significant event in the on-source interval. 
For a given signal model, the GW analysis efficiently recovers signals up to a certain distance that depends on the sensitivity of the detectors at the time and sky position of a given GRB event. 
Knowing the upper limit amplitude of each waveform, $h_{rss}$, the exclusion distance can be computed using the standard siren model as defined in \cite{sutton13}, namely sources with fixed energies emitted in gravitational radiation.
For GRBs under the most optimistic scenarios, the approximate maximum GW emission for CSGs has an energy of about E$_{GW}$ = $10^{-2}$ $M_\odot c^2$, and for ADIs E$_{GW} = 0.1$ $M_\odot c^2$ \cite{abbott2017c}.
%Another possible energy value is E$_{GW}$ = $10^{-8}$ $M_\odot c^2$, which is around the order of GW energy produced in simulations of core-collapse supernovae \cite{christian,sutton13}. 
%value is computed according to \cite{putten} using the relation E$_{GW}$ $\approx$ $0.1M_\odot$$(M/M_\odot)$.

In Table 1 (taken from \cite{abbott2021}) we show the cumulative 90$\%$ confidence level exclusion distances (D$_{90}$) for the sets of waveforms used in the unmodelled search for several emission scenarios.
%with total radiated energy E$_{GW}$ = $10^{-2}$ $M_\odot c^2$.
%(\cite{abbott2021}).
These limits depend on the sensitivity of the detector network and the values obtained for CSGs and ADIs vary roughly over one order of magnitude. 
%For the modeled search - BNS, BBH and BHNS signal- \cite{abbott2021} reported that their D$_{90}$ values are 40$\%$ – 60$\%$ times higher than those reported in \cite{abbott2019b} for the previous LIGO–Virgo observing run. 

\begin{figure}
\centering
\begin{minipage}[c]{.47\textwidth}
  \centering
  \includegraphics[width=1.06\linewidth]{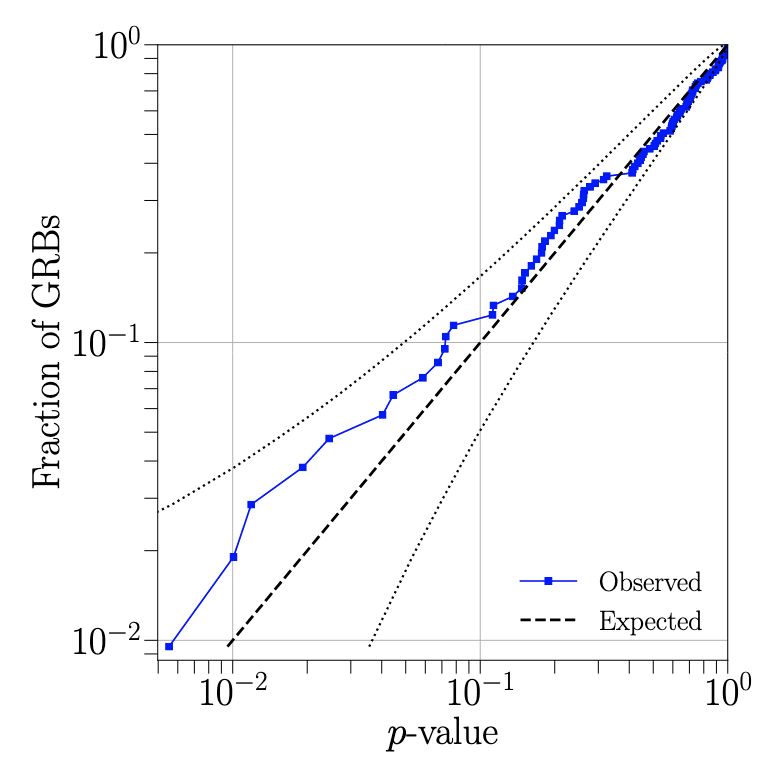}
  \captionof{figure}{The cumulative distribution of p-values. The dashed line represents the
expected distribution under the no-signal hypothesis, with the dotted lines indicating a 2$\sigma$ deviation \cite{abbott2021}.}
  \label{pvalue}
\end{minipage}\hfill
\begin{minipage}[c]{.46\textwidth}
  \centering
  \includegraphics[width=1.01
  \linewidth]{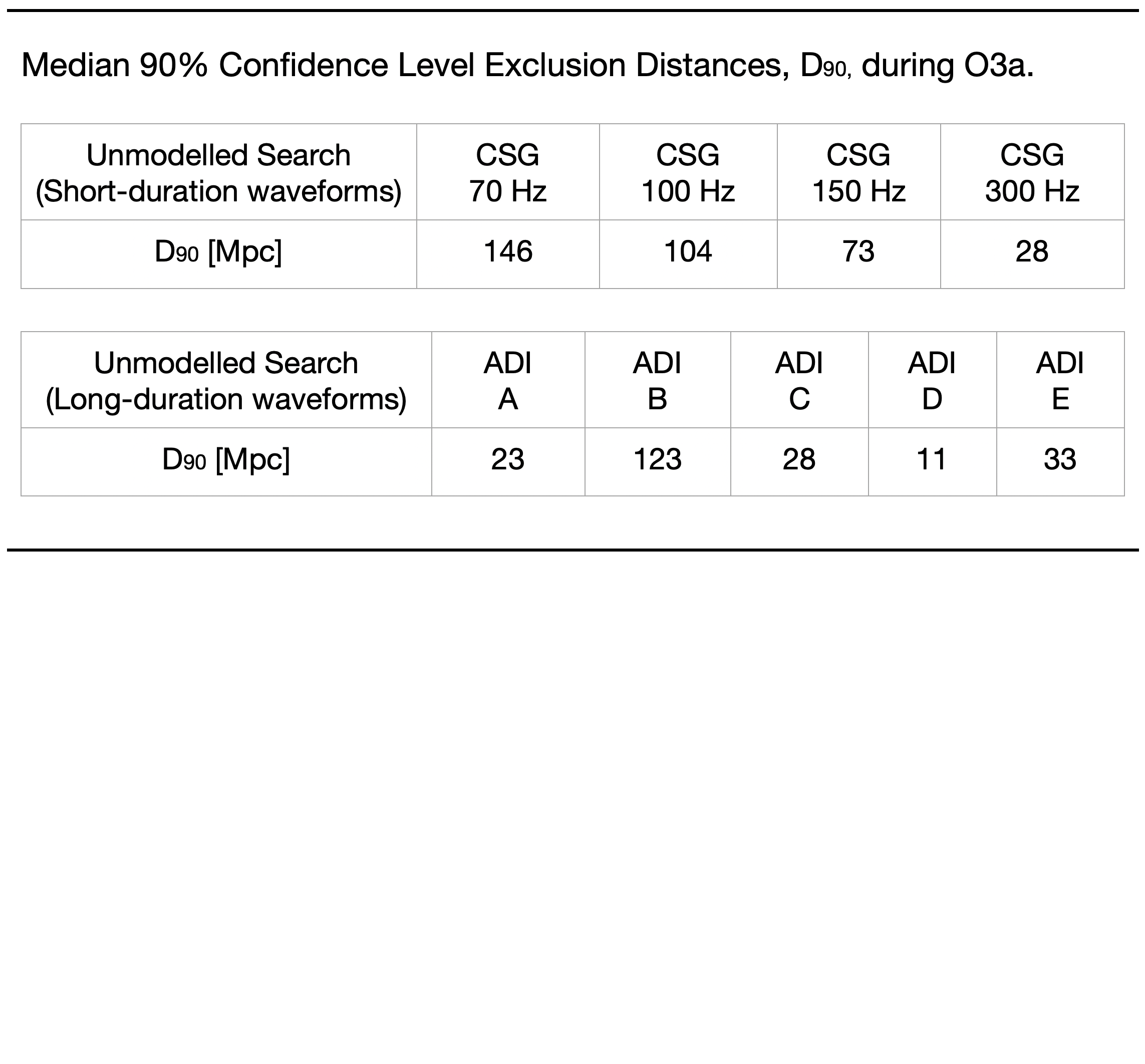}
  \captionof{table}{The median 90$\%$ confidence level exclusion distances during O3a \cite{abbott2021}.}
  \label{table}
\end{minipage}
\end{figure}

\section{Conclusions and future prospects}

We carried out an analysis using the data of LIGO–Virgo observing run O3a to look for GWs coincident with GRBs using an unmodelled search for GW transients. 
From 105 GRBs analyzed with X-pipeline, no GW was detected in coincidence with a GRB, consistently with joint detections of 0.1–1.4 per year during the 2018–2019 observing run of Advanced LIGO and Advanced Virgo \cite{abbott2017b}.
%and 0.3–1.7 per year at design sensitivity  
Having found no coincidence, we set lower bounds on the distances to the progenitors of all analyzed GRBs for different sets of waveforms. 
The D$_{90}$ reported by \cite{abbott2021} includes the largest values published so far for some individual GRBs when considering the unmodelled and modelled searches. Among these, GRB 190610A has a sky localization which includes a nearby galaxy at a luminosity distance of 165 Mpc. 

The search presented here should be viewed as a prototype for future searches for GRBs and other types of coincident phenomena. 
Since much is currently unknown about FRBs, the detection of a GW in coincidence with a FRB could provide insight into the astrophysical origin of these radio transients.
Efforts within the LIGO, Virgo and KAGRA collaborations are being made to analyze data around radio transients detected by instruments operating at frequencies lower than those of the Green Bank or Parkes telescopes \cite{abbott2016b}, such as CHIME \cite{chime}.

This material is based upon work supported by NSF’s LIGO Laboratory which is a major facility fully funded by the National Science Foundation.

\end{document}